\documentclass[12pt]{iopart}

\usepackage{enumerate}
\usepackage{amssymb}
\usepackage{amsbsy}
\usepackage[dvips]{graphicx}
\usepackage{color}
\usepackage{graphics}

\newcommand{\be}{\begin{equation}} \newcommand{\ee}{\end{equation}}
\newcommand{\bea}{\begin{eqnarray}} \newcommand{\eea}{\end{eqnarray}}

\newcommand{\re}[1]{(\ref{#1})}

\newcommand{\pat}{\partial}

\newcommand{\fig}[1]{figure \ref{#1}}
\newcommand{\brt}[1]{[#1]}
\newcommand{\para}{\paragraph}

\renewcommand{\d}{\delta}

\newcommand{\GN}{G_{\mathrm{N}}}

\newcommand{\teq}{t_{\mathrm{eq}}}

\newcommand{\rhom}{\rho_{\mathrm{m}}}

\newcommand{\adot}{\dot{a}}
\newcommand{\addot}{\ddot{a}}

\renewcommand{\H}{\frac{\adot}{a}}
\newcommand{\HH}{\frac{\adot^2}{a^2}}

\newcommand{\av}[1]{\langle{#1}\rangle}
\newcommand{\sQ}{\mathcal{Q}}
\newcommand{\sR}{{^{(3)}R}}

\newcommand{\PRD}[1]{{\it Phys. Rev.} {\bf D#1}}

\newcommand{\APJ}[1]{{\it Astrophys. J.} {\bf #1}}

\renewcommand{\CQG}[1]{{\it Class. Quant. Grav.} {\bf #1}}
\newcommand{\GRG}[1]{{\it Gen. Rel. Grav.} {\bf #1}}

\newcommand{\IJMPD}[1]{{\it Int. J. Mod. Phys.} {\bf D#1}}

\begin{document}

\begin{titlepage}

%\hspace{11cm} March 19, 2008

\title{The effect of structure formation on the expansion of the universe}

\author{Syksy R\"{a}s\"{a}nen}

\address{Universit\'e de Gen\`eve, D\'epartement de Physique Th\'eorique\\
24 quai Ernest-Ansermet, CH-1211 Gen\`eve 4, Switzerland}

\ead{syksy {\it dot} rasanen {\it at} iki {\it dot} fi}

\begin{abstract}

\noindent
Observations of the expansion rate of the universe
at late times disagree by a factor of 1.5--2
with the prediction of homogeneous and isotropic models
based on ordinary matter and gravity.
We discuss how the departure from linearly
perturbed homogeneity and isotropy due to structure
formation could explain this discrepancy.
We evaluate the expansion rate in a dust universe
which contains non-linear structures with a
statistically homogeneous and isotropic distribution.
The expansion rate is found to increase relative
to the exactly homogeneous and isotropic case by
a factor of 1.1--1.3 at some tens of billion of years.
The timescale follows from the cold dark matter
transfer function and the amplitude of primordial
perturbations without additional free parameters.

\end{abstract}

\begin{center}

\noindent {\it This essay was awarded Honorable Mention in the\\2008 Gravity Research Foundation essay competition.}

\end{center}

\end{titlepage}

\paragraph{Three assumptions and a factor of two.}

Cosmological observations show that the early universe
is well described by a model which contains only ordinary
matter (i.e. baryons, leptons, photons, and dark matter),
evolves according to ordinary general relativity and is
exactly homogeneous and isotropic (up to linear perturbations).
However, such a model underpredicts the cosmological distances
measured in the late universe by a factor of about 2.

In a homogeneous and isotropic model, the distance scale is
determined in terms of the expansion rate and spatial
curvature, and the disrepancy can be summarised by saying
that the observed Hubble parameter is a factor of 2 larger
than expected given the matter density
(i.e. $3 H^2\approx4\times8\pi\GN\rhom$ instead of
$3 H^2\approx8\pi\GN\rhom$), or a factor of 1.5 larger given
the age of the universe (i.e. $Ht\approx1$ instead of $Ht=2/3$).
More precisely, the Hubble parameter has fallen more slowly than
predicted, corresponding to acceleration.

Explaining the factor of 2 requires abandoning at least one of
the three assumptions of standard matter, standard gravity and perfect
homogeneity and isotropy. Keeping to homogeneity and isotropy,
it is possible to account for the distance observations by adding
a factor of 3 to the energy density in the form of exotic matter
with negative pressure or introducing repulsive gravity in the same
measure. Such models have two shortcomings.

First, it is difficult to understand why the contributions of
ordinary matter and the repulsive component are roughly equal
today, at around 10 billion years.
This {\it coincidence problem} is somewhat philosophical in nature:
it does not contradict any known physical law or observation.

In contrast, the second problem of homogeneous and isotropic
models is unambiguous: the universe is not perfectly homogeneous
and isotropic (or even perturbatively near homogeneity and isotropy).
There are non-linear structures which are not described by
perturbations around a smooth background, with a distribution
that is statistically homogeneous and isotropic above a scale of
about 100 Mpc \cite{Hogg:2004}.

A universe which is homogeneous and isotropic only statistically
does not in general expand like an exactly homogeneous and isotropic
universe, even on average. This feature of general relativity was
discussed under the name {\it fitting problem} by George Ellis in
1983 \cite{fitting}. However, at the time the observational
situation was not clear enough for factors of order one to
be important. Now that cosmological observations have become
more precise and a discrepancy has arisen, the complication
due to non-linear structures can no longer be neglected.
Also, the fact that structure formation is the most prominent
change in the universe at late times suggests that
it might solve the coincidence problem \cite{Rasanen}.

\para{The Buchert equations.}

The effect of inhomogeneity and/or anisotropy on the average
evolution is called backreaction \cite{Ellis:2005}. Backreaction
is quantified by the Buchert equations, which describe the average
evolution of a rotationless dust space and provide a partial
answer to the fitting problem \cite{Buchert:1999}:
\bea
  \label{Ray} 3 \frac{\addot}{a} &=& - 4 \pi \GN \av{\rho} + \sQ \\
  \label{Ham} 3 \HH &=& 8 \pi \GN \av{\rho} - \frac{1}{2} \av{\sR} - \frac{1}{2} \sQ \\
  \label{cons} && \pat_t \av{\rho} + 3 \H \av{\rho} = 0 \ ,
\eea

\noindent where dot is a derivative with respect to proper time $t$,
$\av{\rho}$ is the average energy density,
$\av{\sR}$ is the average spatial curvature,
$\frac{1}{3}\av{\theta}=\adot/a\equiv H$ is the average
Hubble parameter ($\theta$ is the local volume expansion rate),
and the backreaction variable $\sQ$ contains the
effect of inhomogeneity and anisotropy:
\bea \label{Q}
  \sQ \equiv \frac{2}{3}\left( \av{\theta^2} - \av{\theta}^2 \right) - 2 \av{\sigma^2} \ ,
\eea

\noindent where $\av{\sigma^2}$ is the average shear scalar. The averages
are taken on the hypersurface of constant proper time; for details and
discussion, see \cite{Rasanen:2006, Rasanen:2008}.

If the variance of the expansion rate in \re{Q} is large enough
compared to the shear and the energy density, the average expansion
rate accelerates, as \re{Ray} shows.
The physical reason is simply that the fraction of space in
faster expanding regions grows, so the average expansion rate can rise.

Structure formation is by definition a non-perturbative problem,
and evaluating the expansion rate in a model with realistic
structures is more involved than introducing a new source term
in homogeneous and isotropic models.
It is not feasible to obtain an exact metric.
However, in a statistically homogeneous and isotropic universe,
only statistical information is needed for calculating
the average expansion rate.
The Buchert equations reduce the effect of structures
to the function $\sQ$, which depends only on global statistics.

We may draw an analogy with a classical system of particles.
For a couple of particles, or for small perturbations about a smooth
background, it may be reasonable to look for an exact solution.
However, with many particles and sizeable local
fluctuations, the system can only be treated statistically.
Statistical treatment is also sufficient for evaluating the
interesting properties of such systems, at least when the
coherence length of fluctuations is much smaller than the
scales of interest. In cosmology, practically all observations
are made over scales larger than the homogeneity scale.

\para{The peak model.}

We will evaluate the average expansion rate in a simple model.
We consider a homogeneous and isotropic, spatially flat,
matter-dominated  background with an initial spectrum of
linear Gaussian perturbations.
We follow the evolution of the perturbations into
the non-linear regime statistically.

We model structures as isolated spherical peaks of the initial
density field smoothed on scale $R$, as in the peak model of
structure formation \cite{Bardeen:1986}.
The number density of peaks is determined as a function of
$\delta/\sigma_0(t,R)$, the linear density contrast divided
by the rms density contrast smoothed on scale $R$.
The smoothing scale $R$ is fixed by taking the rms density contrast
to be unity at all times, $\sigma_0(t,R)=1$. The scale $R$ thus
measures the size of typical structures forming at time $t$,
and its growth corresponds to the progress of structure formation
to larger scales.

Since the individual regions are spherical and isolated,
they evolve (in the Newtonian limit) like independent
homogeneous and isotropic universes (see e.g. \cite{Padmanabhan:1993}).
Overdense structures slow down and collapse, underdense voids
expand less slowly and become emptier.

The evolution of the number density is determined by the
power spectrum of linear perturbations. We take a scale
invariant primordial spectrum with the observed amplitude,
and consider two different approximations of the cold dark
matter transfer function.
In addition to the well-known BBKS function \cite{Bardeen:1986},
we use the simple form introduced by Bonvin and Durrer \cite{Bonvin:2005}.
The different results give some indication of the degree
of uncertainty in the results due our simplified modelling assumptions.

The average Hubble rate at time $t$ is given by
\bea
  H(t) &=& \int_{-\infty}^{\infty} \rmd\delta\, v_\d(t) H_\d(t) \ ,
\eea

\noindent where $v_\d(t)$ is the fraction of volume in regions with
linear density contrast $\d$, and $H_\d(t)$ is the Hubble
rate of such regions. The volume fraction is composed of two parts,
$v_\d(t)=s_\d f(\d,t)/( \int_{-\infty}^{\infty} \rmd\delta\, s_\d f(\d,t) )$,
where $s_\d$ is the volume of a region with linear density contrast $\d$
relative to an unperturbed region, and $f(\d,t)$ is the number density
of such regions. See \cite{Rasanen:2008} for details.

\begin{figure}
\hfill
\begin{minipage}[t]{7.5cm} 
\scalebox{1.0}{\includegraphics[angle=0, clip=true, trim=0cm 0cm 0cm 0cm, width=\textwidth]{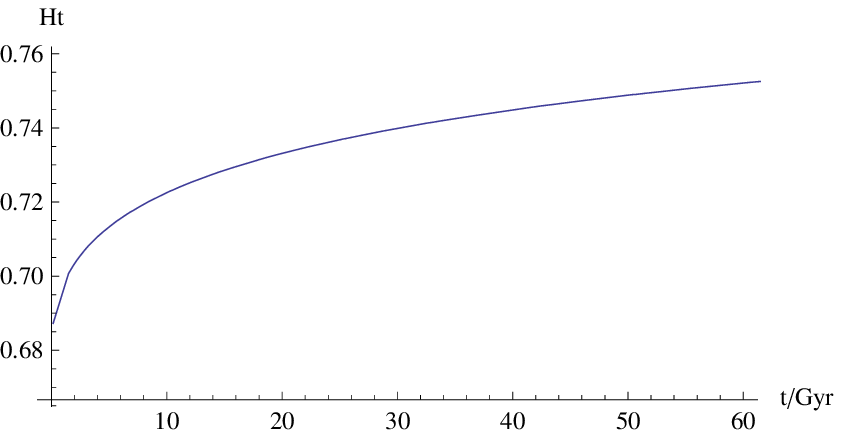}}
\begin{center} {\bf (a)} \end{center}
\end{minipage}
\hfill
\begin{minipage}[t]{7.5cm}
\scalebox{1.0}{\includegraphics[angle=0, clip=true, trim=0cm 0cm 0cm 0cm, width=\textwidth]{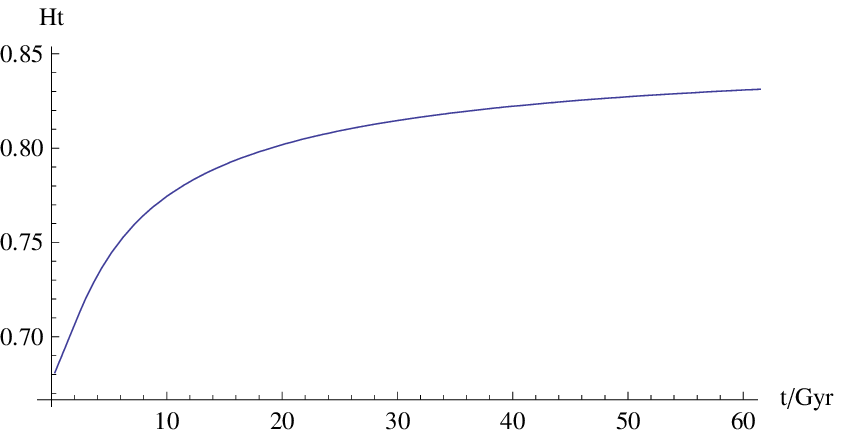}}
\begin{center} {\bf (b)} \end{center}
\end{minipage}
\hfill
\caption{The expansion rate $Ht$ as a function of time for
(a) the BBKS transfer function and (b) the Bonvin and Durrer transfer function.}
\label{fig:Ht}
\end{figure}

Given the initial power spectrum and the transfer function, the evolution
is completely fixed, there are no free parameters to adjust.
The result for $H$ multiplied by $t$ is shown in \fig{fig:Ht}.
In the early universe, the expansion is near
the homogeneous and isotropic case, with $Ht\approx2/3$.
At late times $Ht$ rises, saturating to a final value somewhat
less than unity as voids come to dominate the volume of the universe.
However, the rise is not rapid enough to correspond to acceleration.

The timescale $\approx10^{10}$ years comes from a combination of
the matter-radiation equality time $\teq\approx10^5$ years encoded
in the transfer function and the primordial perturbation
amplitude $\approx10^{-5}$. Because the initial amplitude is
small, it takes long for structures to become important.

\paragraph{Outlook.}

We have demonstrated with a simple model how non-linear structures
lead to an increase in the expansion rate from the homogeneous
and isotropic value.
It is remarkable that the era when the expansion rate increases
comes out roughly correctly without free parameters, showing
how structure formation may solve the coincidence problem.
Given the level of approximation, the fact that the model does
not show acceleration is not particularly worrisome.
In a universe which is not perfectly homogeneous and isotropic,
there is no fundamental difference between acceleration and
deceleration; it is merely a question of how rapidly the faster
expanding regions come to dominate. Acceleration has been explicitly
demonstrated in a dust model with two spherical regions \cite{Rasanen:2006}.

It may be instructive to compare the current situation
to the early years of Newtonian gravity.
Newtonian theory explained the local gravity observations
on Earth, and the two-body solution was very successful in
describing the orbits of the planets.
However, when the two-body solution
was applied to the nearby Earth-Moon system, the result
for the lunar perigee was wrong by a factor of 2.
It was proposed that the inverse square law of gravity
be modified at small distances (of the order of the Earth-Moon
separation) to correct the discrepancy \cite{Gutzwiller:1998}\footnote{In
his book {\it The Structure of Scientific Revolutions}, Thomas Kuhn
uses this as an example of an inadmissible solution to a scientific
puzzle \cite{Kuhn:1996}.}.
However, the solution turned out to lie in the non-linear
aspects of Newtonian gravity: the influence of the Sun cannot
be neglected. Even after a correction of the right order of
magnitude was demonstrated, it took decades before the
non-linear three-body calculation was fully worked out.

Similarly, general relativity has explained the local observations
in the solar system, and the application of the homogeneous
and isotropic solution to the early universe has been very successful.
However, the prediction for the universe nearer to us in time is wrong
by a factor of 2. Whether non-linear structures can explain this
discrepancy is not yet known. However, their effect has to be quantified
before it is possible to draw definite conclusions about other explanations.

\setcounter{secnumdepth}{0}


\section{References}

\begin{thebibliography}{99}

\bibitem{Hogg:2004} Hogg D W \etal,
\newblock {\it Cosmic homogeneity demonstrated with luminous red galaxies}, 2005
\newblock \APJ{642} 54
\newblock \brt{astro-ph/0411197}
%%CITATION = ASTRO-PH 0411197;%%

\bibitem{fitting} Ellis G F R,
\newblock {\it Relativistic cosmology: its nature, aims and problems}, 1984
\newblock The invited papers of the 10th international conference on general relativity and gravitation
\newblock p 215
%%CITATION = NONE;%%
\nonum Ellis G F R and Stoeger W,
\newblock {\it The 'fitting problem' in cosmology}, 1987
\newblock \CQG{4} 1697
%%CITATION = NONE;%%

\bibitem{Rasanen} R\"{a}s\"{a}nen S,
\newblock {\it Dark energy from backreaction}, 2004
\newblock JCAP0402(2004)003
\newblock \brt{astro-ph/0311257}
%%CITATION = ASTRO-PH 0311257;%%
\nonum R\"{a}s\"{a}nen S,
\newblock {\it Backreaction of linear perturbations and dark energy}
\newblock \brt{astro-ph/0407317}
%%CITATION = ASTRO-PH 0407317;%%

\bibitem{Ellis:2005} Ellis G F R and Buchert T,
\newblock{\it The universe seen at different scales}, 2005
\newblock {\it Phys. Lett.} {\bf A347} 38
\newblock \brt{gr-qc/0506106}
%%CITATION = GR-QC 0506106;%%

\bibitem{Buchert:1999} Buchert T,
\newblock {\it On average properties of inhomogeneous fluids in general relativity I: dust cosmologies}, 2000
\newblock \GRG{32} 105
\newblock \brt{gr-qc/9906015}
%%CITATION = GR-QC 9906015;%%N = GRGVA,25,1225;%%

\bibitem{Rasanen:2006} R\"{a}s\"{a}nen S,
\newblock{\it Cosmological acceleration from structure formation}, 2006
\newblock \IJMPD{15} 2141
\newblock \brt{astro-ph/0605632}
%%CITATION = ASTRO-PH 0605632;%%
\nonum R\"{a}s\"{a}nen S,
\newblock {\it Accelerated expansion from structure formation}, 2006
\newblock JCAP11(2006)003
\newblock \brt{astro-ph/0607626}
%%CITATION = ASTRO-PH 0607626;%%

\bibitem{Rasanen:2008} R\"{a}s\"{a}nen S,
\newblock {\it Evaluating backreaction with the peak model of structure formation}, 2008
%\newblock JCAP11(2006)003
\newblock \brt{arXiv:0801.2692v2 [astro-ph]}
%%CITATION = ARXIV:0801.2692;%%

\bibitem{Bardeen:1986} Bardeen J M, Bond J R, Kaiser N and Szalay A S,
\newblock {\it The statistics of peaks of Gaussian random fields}, 1986
\newblock \APJ{304} 15
%%CITATION = ASJOA,304,15;%%

\bibitem{Padmanabhan:1993} Padmanabhan T,
\newblock{\it Structure formation in the universe}, 1993
\newblock Cambridge University Press, Cambridge,
\newblock p 273
%%CITATION = NONE;%%

\bibitem{Bonvin:2005} Bonvin C, Durrer R and Gasparini M A,
\newblock {\it Fluctuations of the luminosity distance}, 2006
\newblock \PRD{73} 023523
\newblock \brt{astro-ph/0511183}
%%CITATION = ASTRO-PH 0511183;%%

\bibitem{Gutzwiller:1998} Gutzwiller M C,
\newblock {\it Moon-Earth-Sun: The oldest three-body problem}, 1998
\newblock {\it Rev. Mod. Phys.}  {\bf 70} 589
%%CITATION = RMPHA,70,589;%%

\bibitem{Kuhn:1996} Kuhn T S,
\newblock {\it The Structure of Scientific Revolutions}, 1996
\newblock 3rd edition
\newblock The University of Chicago Press, Chicago and London,
\newblock p 39, p 81
%%CITATION = NONE;%%

\end{thebibliography}
\end{document}